\documentclass[11pt]{article}
\pdfoutput=1

\usepackage{amsmath,amssymb,amsfonts,graphicx}
\usepackage{ucs}
\usepackage{soul}
\usepackage{epstopdf}
\usepackage{color}
\usepackage[utf8]{inputenc}
\usepackage{amsthm}
\usepackage{fontenc}
\usepackage{tikz}
\RequirePackage{slashed}
\usepackage{cancel}
\usepackage[normalem]{ulem}

\usepackage{color}
\usepackage{latexsym}
\usepackage{amssymb}
\usepackage{booktabs}
\usepackage{comment}
\usepackage{authblk}
\usepackage{float} 
\usepackage{cite}
\numberwithin{equation}{section}

\usepackage[a4paper,top=2cm,bottom=2cm,left=2cm,right=2cm]{geometry}

\usepackage{braket}

 \newcommand{\open}{\sphericalangle}

\newlength\savedwidth

\title{
A constrained fit of the valence transversity distributions from dihadron production}
\date{\today}

\begin{document}

\author[1,2]{J.~Benel}
\affil[1]{\small{ Instituto de F\'isica, Universidad Nacional Aut\'onoma de M\'exico\\
Apartado Postal 20-364, 01000 Ciudad de M\'exico, Mexico} }
\affil[2]{\small{ Universidad Nacional de Ingenier\'ia, Apartado 31139, Lima, Peru} }
\author[1]{A.~Courtoy}

\author[1]{R.~Ferro-Hernandez}

\maketitle

\begin{abstract}
We present a constrained analysis of the valence transversity Parton Distribution Functions from dihadron production in semi-inclusive DIS. While usual extractions of the transversity distributions rely explicitly on the fulfillment of the Soffer bound, the present analysis releases that implicit restriction to implement further explicit constraints through the method of Lagrange multipliers. The results are quantitatively comparable to previous analyses in the kinematical range of data; the qualitative impact of the chosen fitting strategy translates into an increased flexibility in the functional form.
\end{abstract}

\section{Introduction}

Parton Distribution Functions  (PDFs) describe the structure of hadrons in terms of their constituents. At leading twist, the hadronic structure is defined by the distributions corresponding to the vector, axial-vector and tensor Dirac bilinears. The first two, the unpolarized and the helicity Parton Distribution Functions, while  under constant study and improvement, are relatively well-determined. In the last decade, the third leading-twist PDF, the one corresponding to the tensor bilinear~\cite{Artru:1989zv,Jaffe:1991kp,Jaffe:1991ra,Cortes:1991ja}, has been explored in phenomenology. The transversity PDF describes  the distribution of the transverse polarization of quarks  as related to that of their parent hadron. Transversity is a chiral-odd object that is paired, in physical processes, to another chiral-odd object. 

One such candidate object is the {\it single-hadron} Transverse Momentum Dependent Fragmentation Function (TMD FF) depending on the fraction of longitudinal momentum of the fragmenting quark carried by the outgoing hadron, as well as on the trace of the intrinsic transverse momentum of that active quark. The second candidate is the {\it dihadron} FF (DiFF), measured in a fragmentation process into a pair of hadrons whose relative momenta tag the  transverse polarization of the active quark via modulations at the cross-section level. 
Thanks to the data on semi-inclusive pion production in electron-positron annihilation from the $B$ factories, both types of FFs have been measured and used to assess transversity PDFs in either global fits and single-process fits.

Dihadron Fragmentation Functions were proposed by Jaffe {\it et al.}~\cite{Jaffe:1997hf} to access the transversity PDF and successfully implemented in phenomenological studies~\cite{Radici:2001na}, as demonstrated by the recent determinations of the transversity PDF~\cite{Bacchetta:2012ty,Radici:2015mwa,Radici:2018iag}  --that is, the collinear PDF  discussed in this article, as opposed to the Transverse Momentum Dependent (TMD) transversity distributions that have been recently studied in Refs.~\cite{Anselmino:2013vqa,Kang:2015msa,Lin:2017stx}. 

 Yet another possible way to access the transverse polarization of quarks is found through Generalized Parton Distributions relevant for Deeply Virtual Meson Production~\cite{Ahmad:2008hp,Goldstein:2013gra}.

The analyses for the transversity are based on reduced experimental inputs with respecto to the data sets used for the other two leading-twist PDFs. 
All three are subject to analogous first principles, {\it e.g.} positivity constraints and support requirement.  In this paper, we propose to broaden the methodology to incorporate the assessment of first principles directly in the fitting procedure. This methodology palliates the low experimental accuracy of  processes involving transversity PDFs.

To leading-order, the unpolarized, $f_1(x)$, and longitudinally polarized distributions, $g_1(x)$, can be interpreted as probability densities in terms of the helicity representation for Dirac particles --basis of the positivity bound $|g_1^q(x)|<f_1^q(x)$. The interpretation of transversity distribution, $h_1(x)$, must be done in the transversity basis --leading to the bound $|h_1^q(x)|<f_1^q(x)$. Instead, in the helicity basis,  another positivity constraint for $h_1(x)$ has been stated by Soffer~\cite{Soffer:1994ww}
\begin{eqnarray}
|h_1^q(x)|\leq \frac{1}{2}\left(f_1^q(x)+g_1^q(x)\right)\quad,
\label{eq:SB}
\end{eqnarray}
for each flavor of quark and anti-quarks separately. If the Soffer bound is realized at a low factorization scale $Q$, it is preserved under QCD evolution towards higher scales~\cite{Goldstein:1995ek,Vogelsang:1997ak}. The bound in Eq.~(\ref{eq:SB}) has been treated as a first-principle constraint in all extractions of transversities until now. Together with the support, $x\in [0,1]$, they constitute the basic properties that any parametrization of the transversity PDF must satisfy. The implementation of positivity constraints generates a hierarchical dependence on the unpolarized PDF for the helicity PDF and both the unpolarized and helicity PDFs for the transversity.

As a result of the fundamental characteristic of the PDFs to describe the internal structure of hadrons in a given spin configuration, their first Mellin moments embody the fundamental charges when originated from conserved currents.  The first moment of the transversity distribution is known as the tensor charge, in analogy to the vector and axial charge. The tensor charge provides interesting information for processes involving hadrons that are also relevant for searches of New Physics~\cite{Bhattacharya:2011qm,Courtoy:2015haa,Gao:2017ade,Yamanaka:2017mef}. 

On the one hand, recent analyses of the transversity highlight a disagreement between the data and the expression of the Soffer bound in some kinematical regions~\cite{Bacchetta:2012ty,Kang:2015msa,Radici:2018iag}. On the other hand, the tensor charge for the valence and isovector flavor combinations as determined through phenomenology and lattice~\cite{Yamanaka:2018uud,Alexandrou:2019brg,Hasan:2019noy,Harris:2019bih} differ, with the largest discrepancy coming from the $u$-valence contribution.
In light of the phenomenological conflicts with the Soffer bound and the latest lattice QCD determinations of the tensor charge, we analyze the statistical probability of the {\it expression of} the Soffer bound given the semi-inclusive DIS data. Subsequently this statistical significance will be included in the minimization procedure, leveraging the choice of the parametrization. The latter will be chosen, first,  to fulfill the support in Bjorken $x$ and, second, so as to optimize the assessment of the uncertainty that reflects as objectively as possible the error coming from both the experimental data and the theory, {\it i.e.} the implementation of the positivity bounds here.
 The methodology includes a further iteration of the minimization imposing inequality constraints to ensure a smooth behavior for $x\to 1$ through the method of Lagrange multipliers. We carry out the new analysis based on the point-by-point extraction for proton and deuteron SIDIS data of Ref.~\cite{Radici:2015mwa}. There have been no recent improvements from the theory side or new determinations of DiFFs that would justify a complete reanalysis of the two-hadron SIDIS asymmetry.

This paper is organized as follows. 
Our methodology is described in Section~\ref{sec:metho}. 
In Section~\ref{sec:res} we discuss our results --with a particular emphasis on the tensor charge, and then conclusions and possible extensions of this work are drawn in Section~\ref{sec:conclu}. The state-of-the-art formalism  related to dihadron production in semi-inclusive DIS is overviewed in Appendix~\ref{sec:form}.

\section{Methodology}
\label{sec:metho}

The determination of the valence transversities is obtained in three main steps. 
We work within  the hypothesis that first principles constraints can be implemented through the fitting procedure to optimize the collective information of the $N_d=22$ data points and the first principles. We first assess the goodness of the Soffer bound given the data using a Bayesian approach. This first step allows a reweighting of each data point, taking into account the Soffer bound implicitly. Then the fulfillment of other theoretically justified behaviors is achieved through a constrained fit using the  method of Lagrange multipliers. 
Finally, the constrained fit is repeated $N$ times using the bootstrap technique as an integral part of the error treatment~\cite{Radici:2015mwa}.

\subsection{Probability of the Soffer Bound}

The practical implementation of the Soffer bound relies on the other two leading-twist PDFs, to be precise on the fits of the unpolarized and helicity PDFs. An uncertainty is attached to each PDF value at the corresponding $(x, Q^2)$-bin that should combine the proper uncertainty --output of the fitting procedure, and an error coming from the various choices for physical parameters and hypothesis, {\it e.g.} the value of $\alpha_s(M_Z^2)$ or the order in perturbation theory.

In the standard parameterizations of $h_1(x)$, the conservation of the bound is ensured by construction at the level of the functional form at $Q_0$ and is conservatively fulfilled under QCD evolution~\cite{Vogelsang:1997ak}. Hence, a theoretical error could analogously be attributed to the Soffer bound (SB) in the first place, introducing an extra flexibility on the parametrization. 

On the other hand, the confrontation of the SB with the combination of transversities extracted from proton and deuteron targets, Eqs.~(\ref{eq:prot}-\ref{eq:deut}), also depends on the approximations used in the extraction of the DiFFs. The mentioned theoretical error should ideally comprise both sources of uncertainty.
Thus,  we would like to estimate the goodness of the Soffer bound --expressed in terms of PDF parameterizations-- given the transversity combinations extracted from HERMES~\cite{Airapetian:2008sk} and COMPASS~\cite{Braun:2015baa,Braunthesis} data and  implicitly include the bound as a source of uncertainty in the fitting procedure.
\\

\begin{figure}[t]
\begin{center}
		\includegraphics[scale= .65]{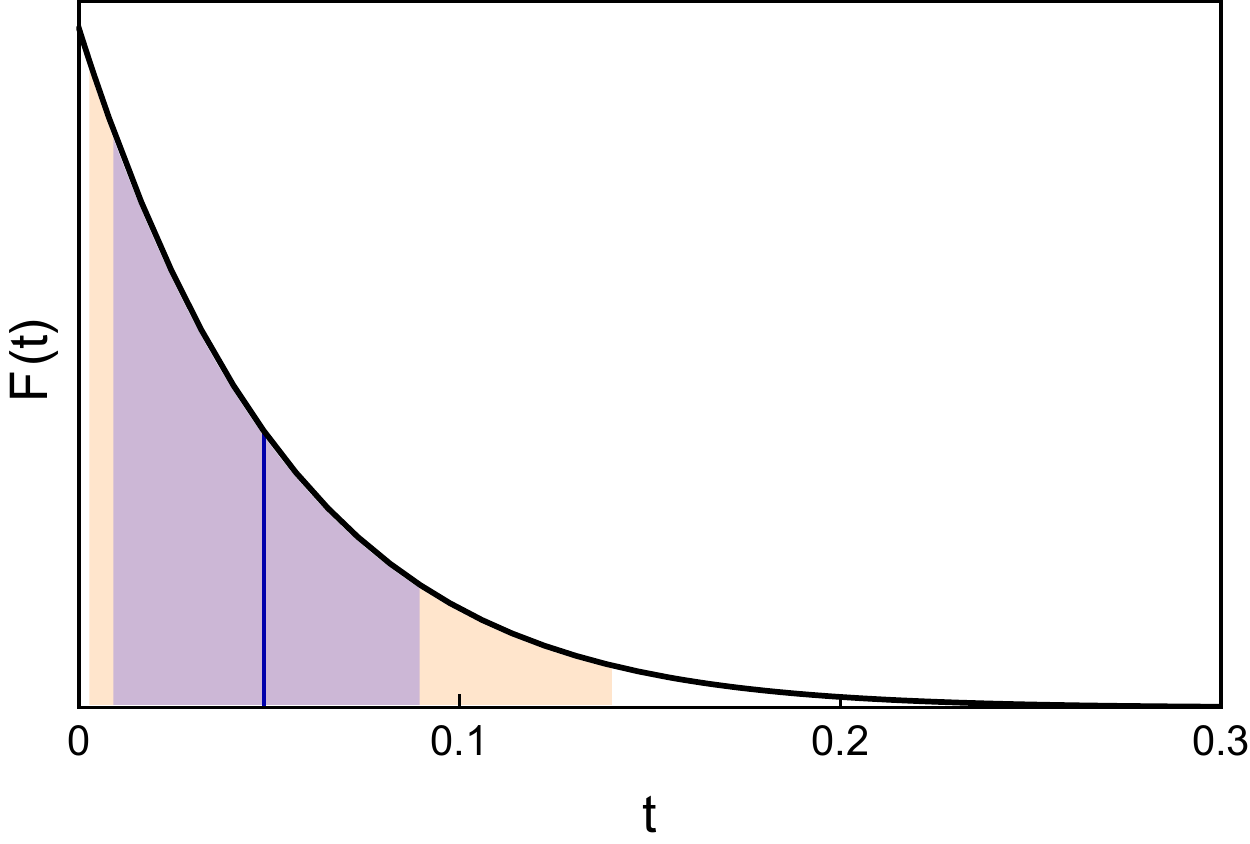}
\end{center}
\caption{Distribution of the hyperparameter $t$ as given by Eq.~(\ref{eq:t_distri}). The vertical blue line corresponds to the central value ${\bar t}$, the blue shaded area to the $68\%$ CL evaluated from the upper/lower bound of the integral of $F(t)$. The orange shaded area corresponds to the $95\%$ CL.  The $y$-axis has arbitrary units.}
\label{fig:t_distri}
\end{figure}
%

For that purpose we evaluate the probability of the transversity PDF lying outside the SB. We first map each experimental
point, which has a Gaussian distribution, into an in-out case. 
Suppose that we know the limits of the bound exactly, {\it i.e.}, we can identified unequivocally whether the data are inside or outside the bound:  $\theta_{j}$ represents that probability, %
\begin{equation}
p\left(\theta_{j}|A_{j}\right)=\begin{cases}
\delta\left(\theta_{j}-1\right) & \mbox{for}\quad A_{j}\in \mbox{region}\\
\delta\left(\theta_{j}\right) & \mbox{for}\quad A_{j}\notin \mbox{region}
\end{cases}\quad,
\label{eq:thetaA}
\end{equation}
where $A_{j}$ is the true value of the transversity combinations of Eqs.~(\ref{eq:prot}-\ref{eq:deut}) evaluated for~$\nolinebreak[1]{j=1,...,N_d}$. The region is the area comprised by the inequality of Eq.~(\ref{eq:SB}) evaluated in each corresponding kinematical bin  using MSTW08LO  at LO~\cite{Martin:2009iq} for $f_1(x_j, Q_j^2)$ and JAM15~\cite{Sato:2016tuz}, at NLO, for $g_1(x_j, Q_j^2)$. 

We define a hyperparameter $t$ corresponding to the probability that the data value lies outside the bound,
\begin{equation}
p\left(\theta_{j}|t\right)=\left(1-t\right)\delta\left(\theta_{j}-1\right)+t\delta\left(\theta_{j}\right)\quad,
\end{equation}
namely a prior distribution for $\theta_{j}$ given $t$. The true values for the transversity combinations can only be inferred through the actual data, such that 
\begin{equation}
p\left(\theta_{j}|\mbox{data}\right)=\int p\left(\theta_{j}|A_{j}\right)p\left(A_{j}|\mbox{data}\right)dA_{j}\quad,
\end{equation}
where the probability of $A_{j}$ given the data evaluates the distance of the data point $j$ to the bound, considering both the uncertainty on the data and on the bound. The distributions are assumed to be Gaussians centered on the experimental value, for the data, or as explained after Eq.~(\ref{eq:thetaA}), for the bound, with the corresponding standard deviation. We assume an error from going from NLO to LO for the helicity PDF based on the NLO-LO difference for unpolarized PDFs. 

The final desired probability is expressed, using Bayes theorem,  as
\begin{eqnarray}
p\left(t|\mbox{data}\right) = N\, \pi(t)\int \, \displaystyle\prod_{j=1}^{N_d}p\left(\theta_{j}|t\right) \, p\left(\theta_{j}|\mbox{data}\right)\, d\theta_{j}\equiv F(t)\quad,
\label{eq:t_distri}
\end{eqnarray}
$N=\int_0^1\, F(t) \, dt$ is the norm, $\pi(t)$ is the prior for $t$ that is chosen to be flat, and the probabilities are defined above.  Evaluating $F(t)$, we find  a distribution with a mode at $t=0$ and a central value for $t$,%
\begin{eqnarray}
{\bar t}&=&\int_0^1\, t\,F(t) \, dt=0.049\pm 0.040\quad,
\label{eq:tave}
\end{eqnarray}
for the  $68\%$ confidence interval. It is illustrated in Fig.~\ref{fig:t_distri}.

\begin{figure}[t]
\begin{center}
		\includegraphics[scale= .45]{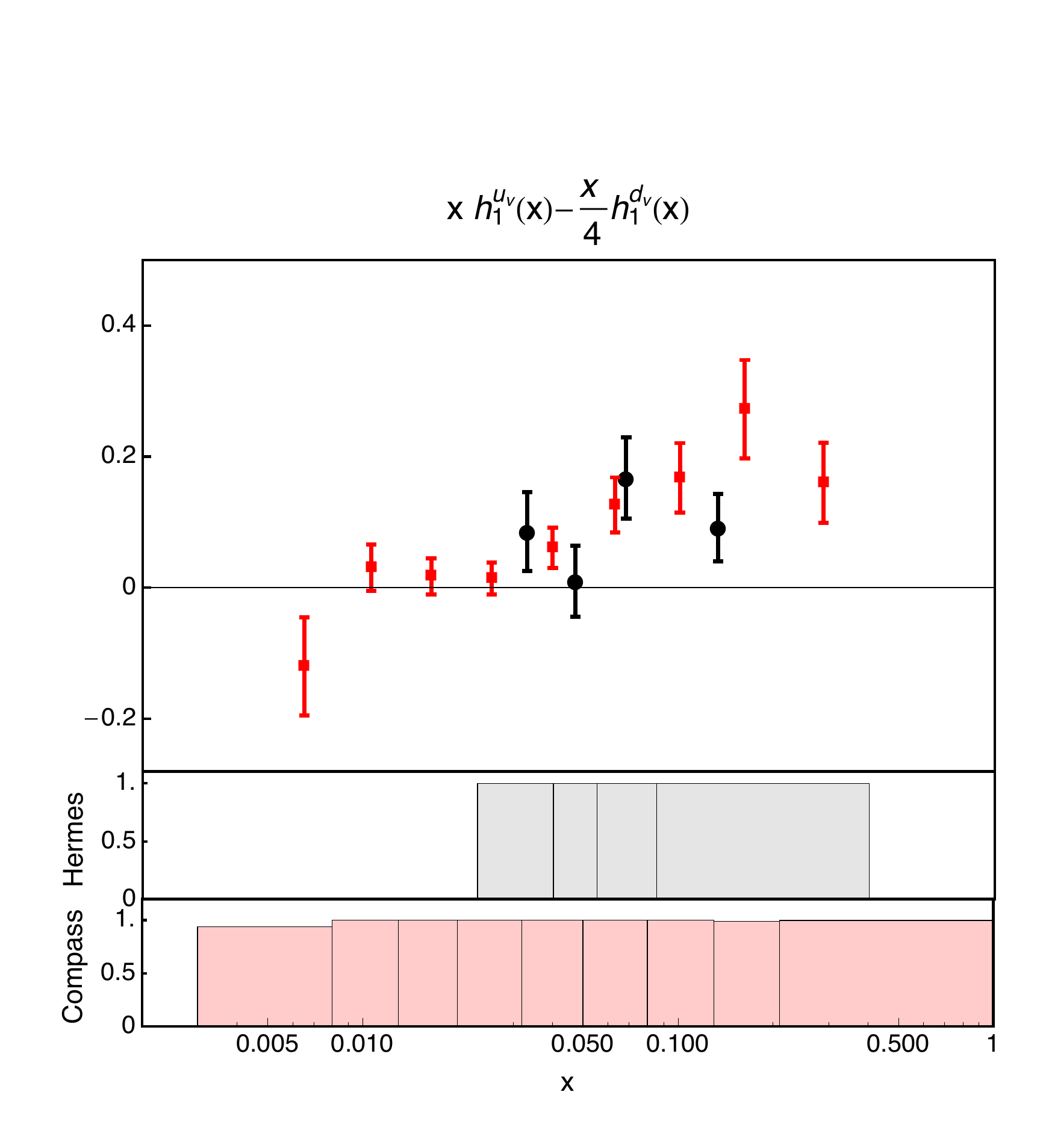}
		\includegraphics[scale= .45]{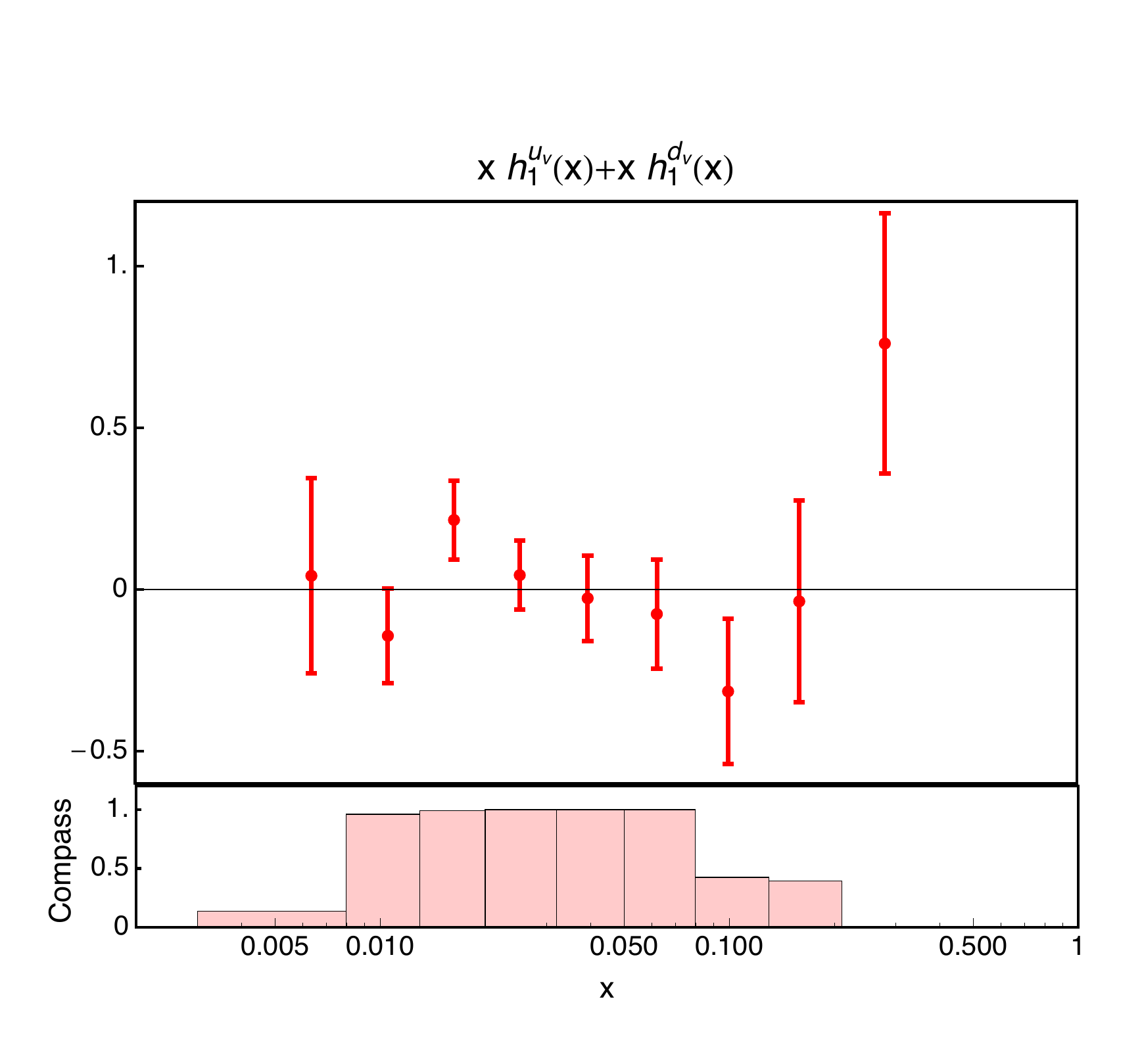}
\end{center}
\caption{Combinations of transversities for the proton (left) and the deuteron (right) as extracted in Ref.~\cite{Radici:2015mwa}. The bottom plots show the statistical weight for each bin as evaluated through $F(t)$.  }
\label{fig:reweighted}
\end{figure}

How should we interpret the result of Eq.~(\ref{eq:tave})? It is highly unlikely  that the bound is incorrect by more than 10\%. 
Still there exists a window through which the agreement of the implemented SB and the transversity combinations is poorer. Since the SB is introduced here as first principle, we translate this result  into a relaxation of the expression of the bound through a Bayesian reweighting bin-by-bin in the objective function --{\it i.e.} here the $\chi^2$. The weight is obtained as follows,
\begin{eqnarray}
w_j&=&\int_{1-p\left(A_{j}|\mbox{\scriptsize data}\right)}^1\, F(t) \, dt\quad.
\end{eqnarray}
In Fig.~\ref{fig:reweighted}, the statistical weight is represented for each bin in the bottom plots. It can be appreciated that the proton combination is almost not affected by this procedure while the lowest and highest $x$-bins of the deuteron combination happen to statistically lie inside that window of poor agreement, as expected from previous analyses~\cite{Bacchetta:2012ty,Radici:2015mwa}.
\\

In the bootstrap technique, the minimization of the objective function is performed $N=200$ times. As explained in, {\it e.g.}, Ref.~\cite{Radici:2015mwa}, $N$ replicas of the extracted combinations $\left(x\, h_1^{p/D}\right)_j$, with $\nolinebreak[1]{j=1,...,N_d}$, of Eqs.~(\ref{eq:prot}-\ref{eq:deut}) are generated randomly within the  data $1\sigma$ error bars. As reweighting the overall chi-square function can be understood as  a scaling of the error,
\begin{eqnarray}
 \sigma_j^2\to \sigma_j^2/w_j\,,\qquad  j=1,...,N_d\,
 \end{eqnarray}
  the $N$ replicas are generated within the corresponding extended Gaussian error bars.   The objective function can be written, for each replica $r$, as 
\begin{eqnarray}
  \chi^2_r\left(\{p^{I}\}\right)&=&\sum_j \,w_j \frac{\left[x_jh_{1\,\mbox{\tiny theo}}^{p/D}\left(x_j ; \{p^{I}\}\right)- \left(xh_{1}^{p/D}(x)\right)_{j,\,r}\right]^2}{\sigma_j^2}\quad,
  \label{eq:chi2main}
\end{eqnarray}
where $\{p^{I}\}$ is the set of free parameters to be determined for each replica $r$. In the next Section, the function $h_{1\,\mbox{\tiny theo}}^{p/D}$ will be extensively described.

The same exercise has been repeated with CT09MC1~\cite{Lai:2009ne} in both Eqs.~(\ref{eq:prot}-\ref{eq:deut}) and  the expression of the SB. The lightest weights for the deuteron combination slightly change, repercussions will be discussed in Section~\ref{sec:res}. The overall conclusions of this section are not affected by the choice of LO unpolarized PDF.

\subsection{Parameterization and Constraints}

Now that  the main first principle based constraint has been implicitly included through a statistical reweighting --which is effectively implemented at the stage of bootstrapping the data-- the parameterization can take a more adaptable form: the data alone lead the determination of the free parameters of the chosen form. An unbiased parametrization is particularly welcome in kinematical ranges with little to no data or, a fortiori, kinematical ranges that exhibit apparent conflict with the positivity bounds.  The latter are best accommodated through a flexible form that could realistically be contrasted with future data.
We wish to impose the required support through the functional form. It is indeed possible for the up distribution. However, the behavior of  $x\,h_1^{d_v}$ is affected by bigger errors  as it is dominated by the deuteron data, and a judicious choice of parametrization is not sufficient to ensure the behavior at the end-point $x=1$. A smooth fall-off in $x$ is expected by the underlying QCD evolution occurring at higher $x$-values --see {\it e.g.} \cite{Scopetta:1997qg}. This observation directly relates to the power-law fall-off ruling the two other leading-twist PDFs.   We tame undesired behaviors in the large-$x$ region for the down distribution through a constrained fit using the method of Lagrange multipliers --see {\it e.g.}~\cite{Behnke:2015}.

\begin{figure}[t]
\begin{center}
		\includegraphics[scale= .53]{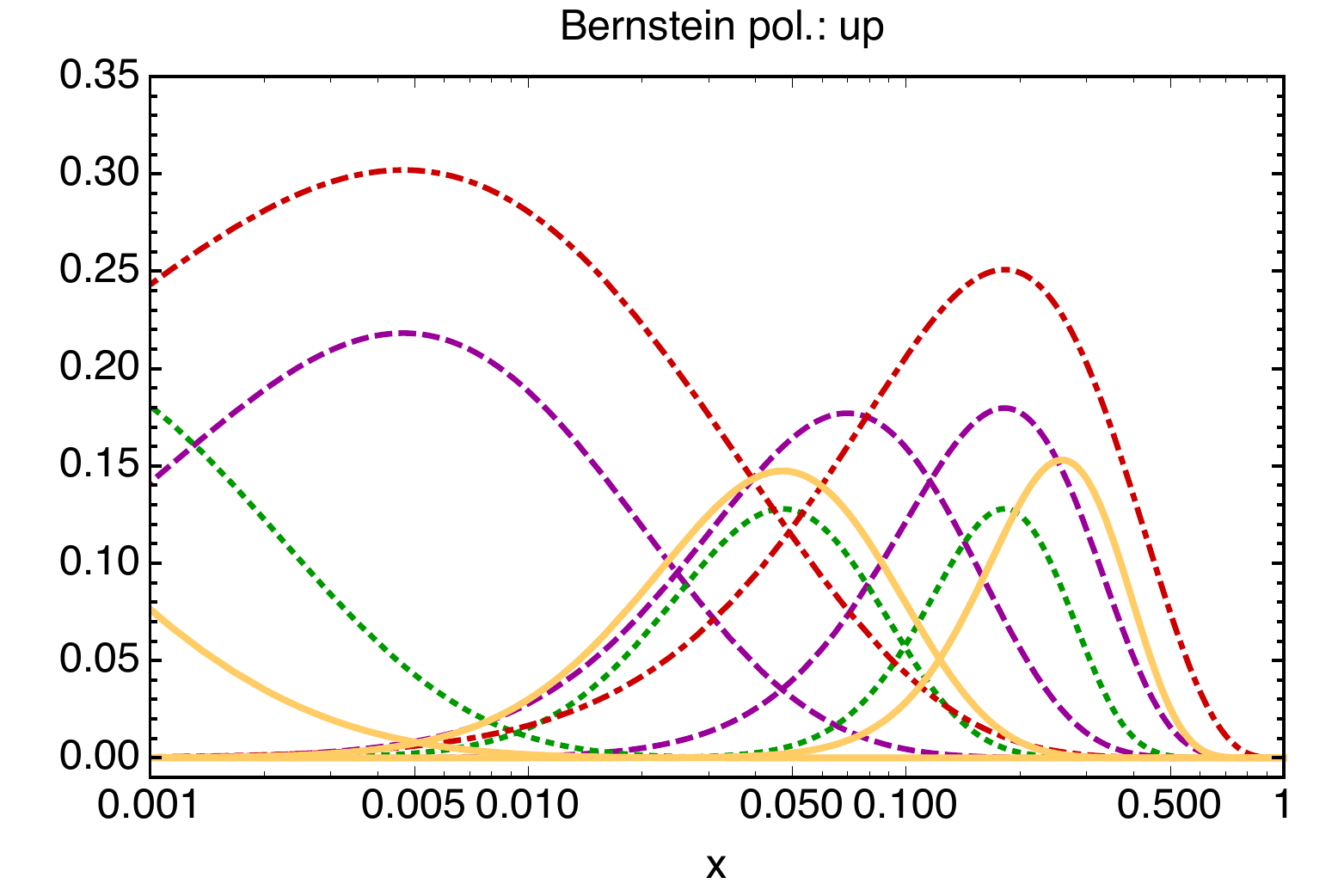}
		\includegraphics[scale= .53]{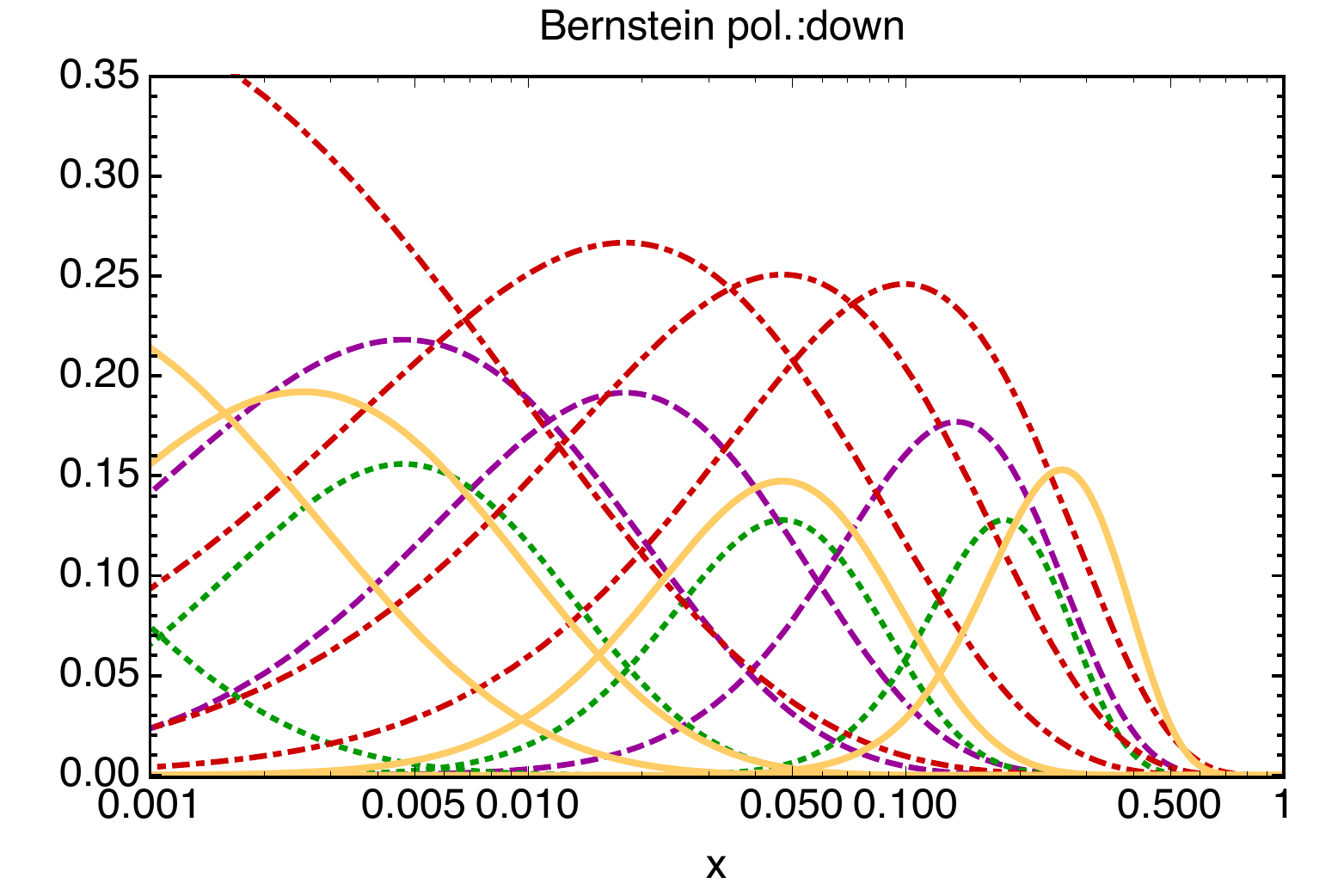}
\end{center}
\caption{Bernstein polynomials $B_{k, n_i}\left(g(x)\right)$ used in the functional form for the valence up transversity (left) and the valence down transversity (right). The degree of the polynomials is, respectively, $n=~\{10, 20, 30, 40\}$ in red with dot-dashed contours, purple/dashed, yellow/full and green/dotted. See text.}
\label{fig:BPUD_44}
\end{figure}

The minimization itself contains two steps. First, the objective function  is minimized via a non-linear least-square, determining the set of best fit parameters, $\{p^I\}$. The  definition of the $ \chi^2(\{p^I\})$, Eq.~(\ref{eq:chi2main}), requires  a functional form for $h_{1\,\mbox{\tiny theo}}^{p/D}(x)$.
The first considerations while determining the parameterization is to guarantee integrability and support at $x=0$,
 \begin{eqnarray}
x\,h_1^{q_v}(x)&=&x^{1.25}\times P_n^q\left(g(x)\right)\quad,
\label{eq:1stFF}
 \end{eqnarray}
 where the exponent has been chosen based on previous outputs and the expected small-$x$ behavior~\cite{Kovchegov:2018zeq} to be slightly modified by the polynomial in $x$. The latter, $P_n^q\left(g(x)\right)$, of order $n$,  can be as flexible as the data allow for. 
 We choose to express $P_n$ in terms of Bernstein polynomials as has been done in Ref.~\cite{Dulat:2015mca}. 
 They are defined as
  \begin{eqnarray}
  B_{k, n}(x)&=& \left(\begin{array}{l}n\\ k \end{array}\right) x^k (1-x)^{n-k} \quad,
   \end{eqnarray}
 with $\left(\begin{array}{l}n\\ k \end{array}\right) $ the binomial coefficients.
 Those polynomials have the advantage of selecting particular regions in $g(x)$ and as such can be employed to emphasize particularly  relevant kinematical regions. That is, for semi-inclusive DIS, the low-$x$ region becomes relevant with respect to the valence and large-$x$ region. A rescaling of the variable will help the parameterization adjusting the data, we choose $g(x)=x^{0.3}$. A desirable feature of the polynomials in Eq.~(\ref{eq:1stFF}) is a statistically representative error outside the data range, yet in agreement with the first principle constraints at hand.  In order to span the Bjorken variable range as significantly as possible, we use four different degrees for the polynomial $P_n$ --thus  using four different functional forms distinguishable by their order $n$-- and optimize the number of Bernstein polynomials by trial and error. In Fig.~\ref{fig:BPUD_44}, we show the polynomials that we have adopted, respectively, for the up and the down parameterization. The functional form is now  expressed as 
 \begin{eqnarray}
x\,h_{1,i}^{q_v}\left(x ; p_{i,k}^q\right)&=&x^{1.25}\,\sum_{k=\{\kappa_{q,i}\}} \,p_{i,k}^q\, B_{k, n_i}\left(g(x)\right)\quad,
\label{eq:finalFF}
 \end{eqnarray}
 with $i=1,\cdots,4$, $n=\{10, 20, 30, 40\}$ and where the $p_{i,k}^q$ are the free parameters which number varies for each $i$. The set of values $\{\kappa_{q,i}\}$ has length $n_{\kappa_{u,i}}=\{2,3,3,3\}$ and $n_{\kappa_{d,i}}=\{4,3,4,3\}$ and their values have been chosen to flexibly cover the relevant region consistently with the data, as can be seen in Fig.~\ref{fig:BPUD_44}. The four functional forms are shared among the $N$ replicas, each $x\,h_{1,i}^{q_v}$ will be evaluated $N/4$ times.
 \\

 In the following step, we will clarify our choice for a limited coverage of the functional form for values of $x\gtrsim 0.6$ for the up and $x\gtrsim 0.5$ for the down.  As mentioned above, we need to ensure a smooth fall-off of the transversity in the limit $x\to 1$ that, for the $d$ contribution, cannot be achieved exclusively  from the choice of functional form. In most cases, a second step will be required to constrain the functional form in an allowed region.
 When the objective function is subject to  $m$ constraints of the form $C_l(\{p'\})=0$ with $l=1,\cdots m$, the later are imposed through the Lagrangian
\begin{eqnarray}
{\cal L}(\{p'\},\{\lambda\})&=&  \chi^2(\{p'\})+\sum_{l}^m \lambda_l C_l(\{p'\})\quad,
\label{eq:lagr_param}
\end{eqnarray}
to which a stationary point of ${\cal L}$ is found minimizing with respect to the parameters $\{p'\}$ and the Lagrange parameters $\{\lambda\}$. 

We define a new objective function that depends on the new set of best fit parameters, $\{p^{II}\}$, which consists in the set made of $p^{q\,II}_{i,k}$, and replaces the set obtained through the first minimization, $\{p^{I}\}$. 
We guide the large-$x$ behavior of the down parameterization only, using the following $N_c=4$ constraints
\begin{eqnarray}
C_{i,l}^{d_v}\left(p^{d\,II}_{i,k}\right)&=&x_l\,h_{1,i}^{d_v}\left(x_l ; p^{d\,II}_{i,k}\right)< \epsilon_l\quad\;\; \mbox{for} \quad l=1,\cdots,N_c/2\quad, \nonumber\\
C_{i,l}^{d_v}\left(p^{d\,II}_{i,k}\right)&=&x_l\,h_{1,i}^{d_v}\left(x_l ; p^{d\,II}_{i,k}\right)>- \epsilon_l\quad \mbox{for} \quad l=1,\cdots,N_c/2\quad,
\label{eq:ineq}
\end{eqnarray}
with $x_l=\{0.3, 0.55\}$ and  $\epsilon_l=\{0.2, 0.1\}$.  In other words, we add 4 degrees of freedom to our problem. The values for $\epsilon_l$ have been set considering the steepness of the functional forms and the trend of $f_1(x,Q^2)$ and $g_1(x,Q^2)$ through which is emulated the  shift to small values of $x$ induced by DGLAP. 

 In previous --unpolarized and longitudinally polarized-- PDF determinations,  the method of the Lagrange multipliers has been made popular for error estimation~\cite{Pumplin:2000vx}. In the present approach, this method is used to impose limits on the fit parameters. This last step completes a methodology that focuses on the adaptability of the parametrization to constraints, in opposition to a parametrization that is constrained a priori.

\begin{figure}[t]
\begin{center}
		\includegraphics[scale= .53]{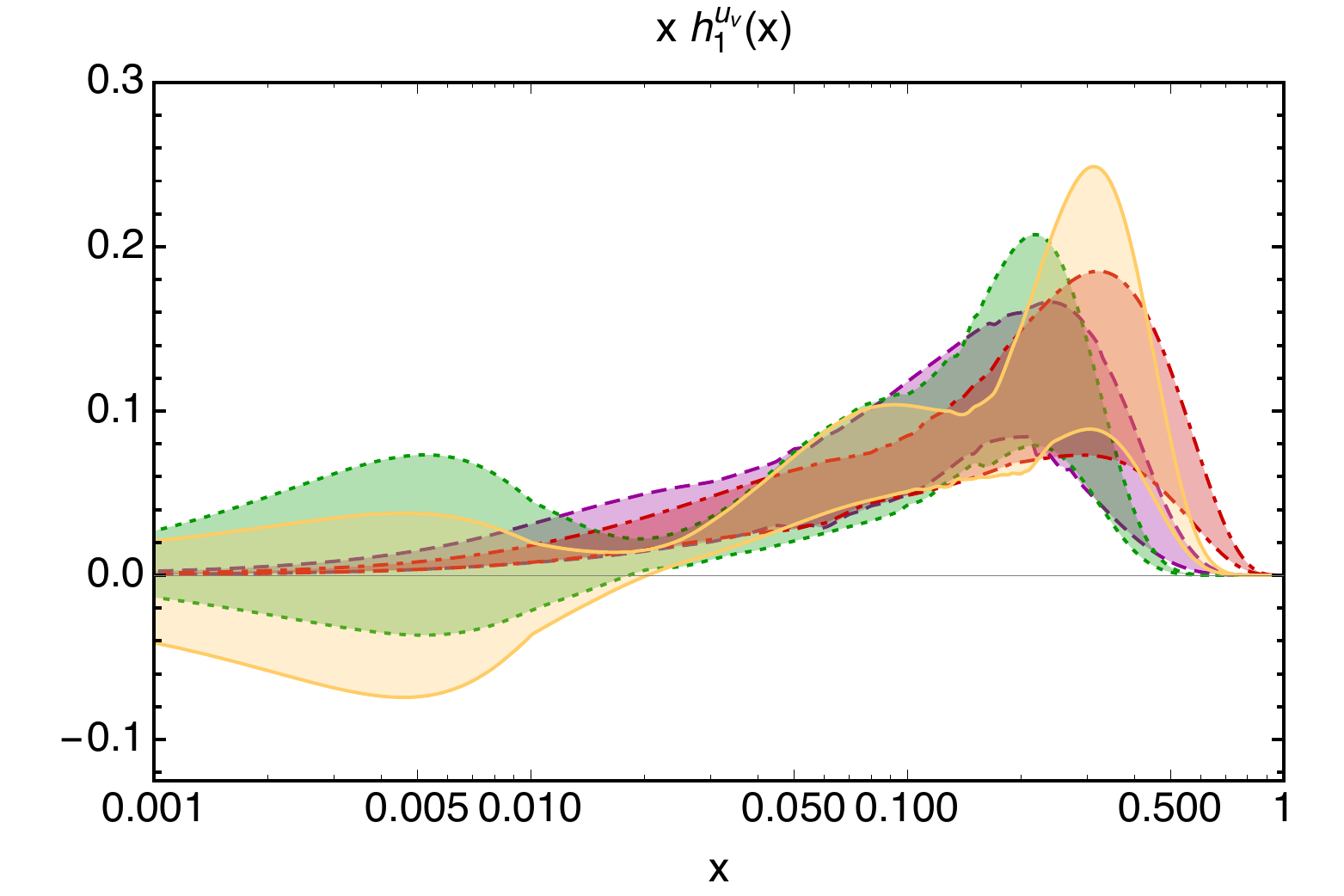}
		\includegraphics[scale= .53]{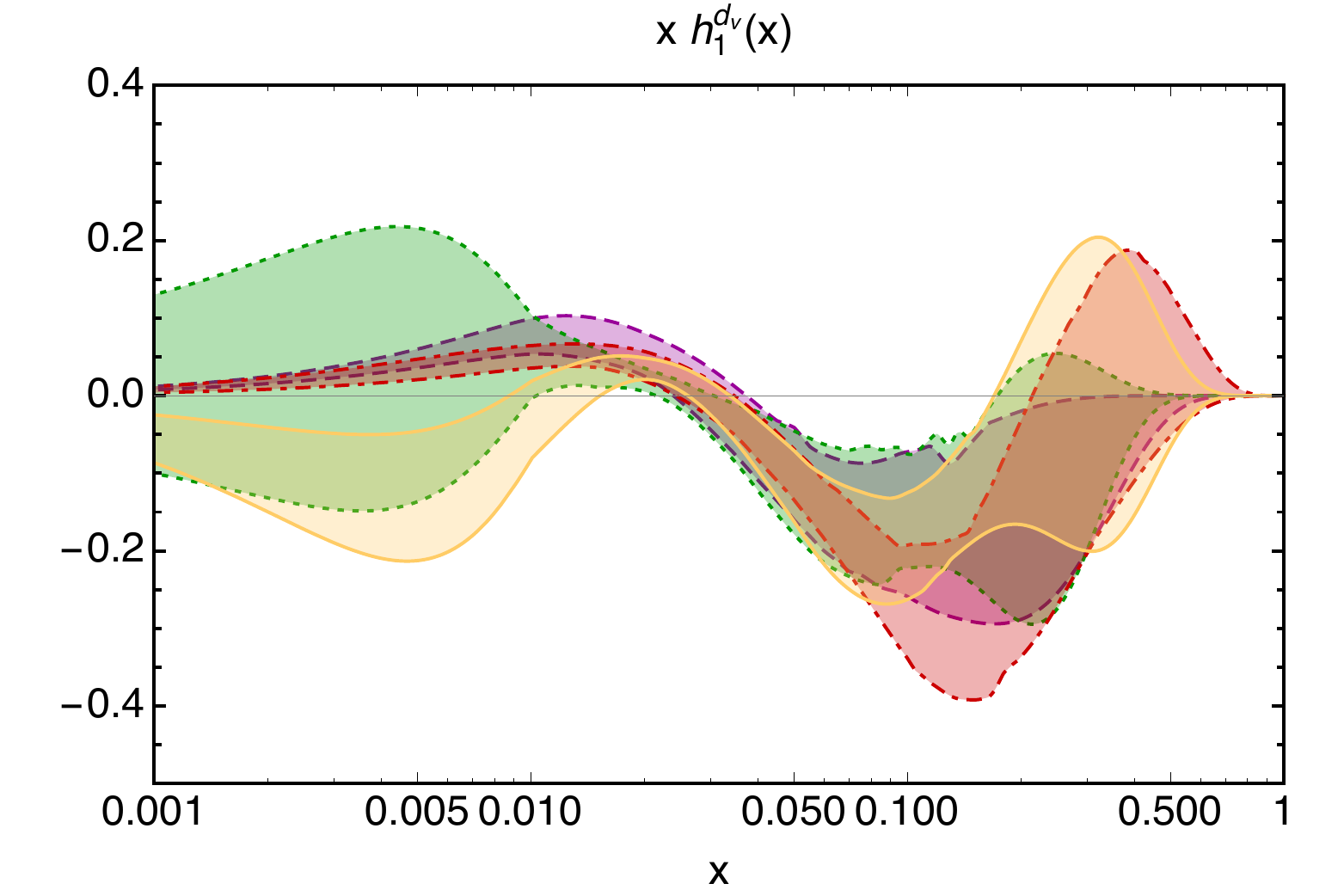}
\end{center}
\caption{Valence transversities:  $68\%$ envelope for the present fit for the four different degrees of the Bernstein polynomial of the functional form, respectively, $n=10, 20, 30, 40$ in red with dot-dashed contours, purple/dashed, yellow/full and green/dotted, for the up  (left) and the down (right).}
\label{fig:UD_44}
\end{figure}

\begin{figure}[t]
\begin{center}
		\includegraphics[scale= .53]{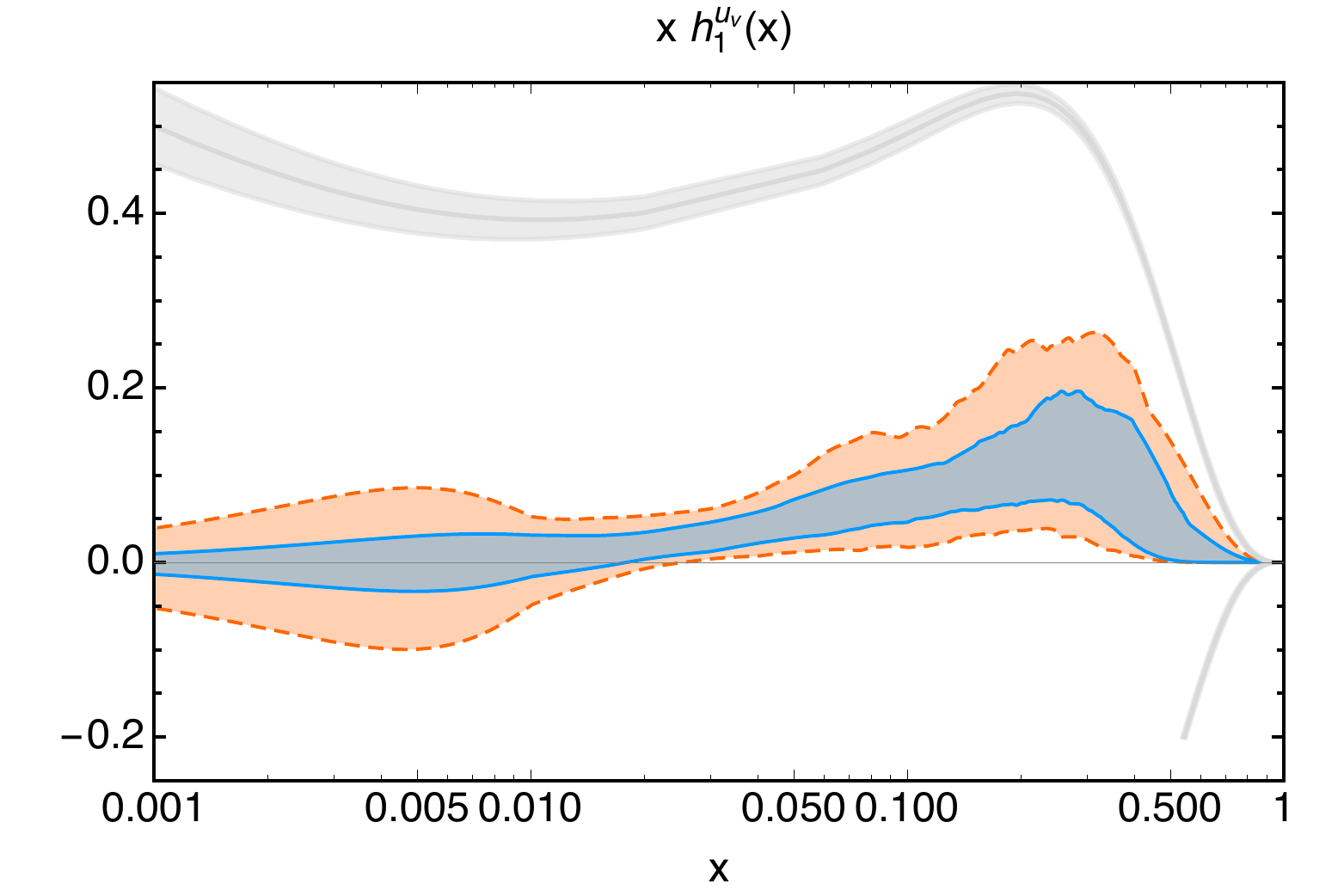}
		\includegraphics[scale= .53]{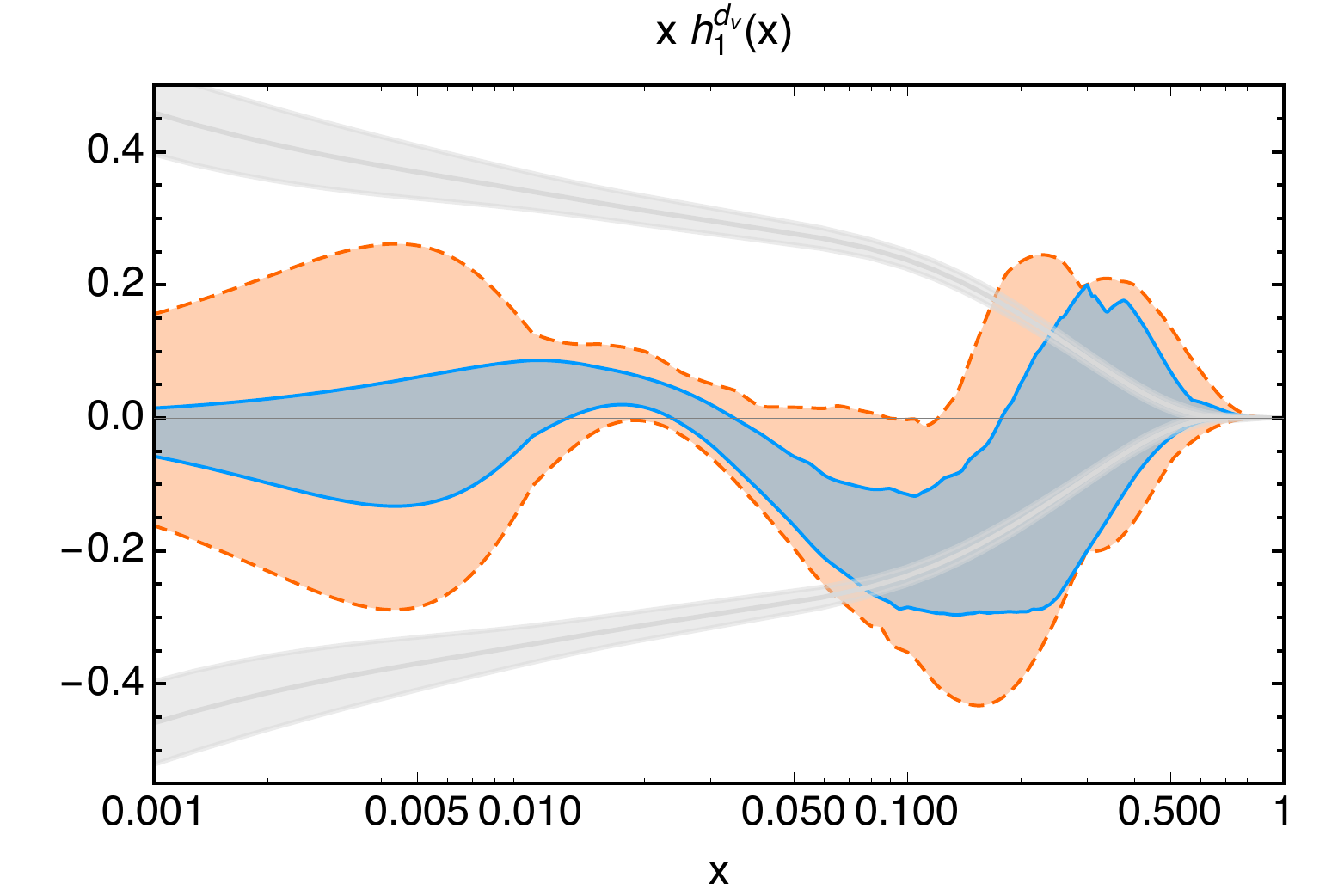}
\end{center}
\caption{Valence transversities for the up (left) and the down (right) at $68\%$ in blue (with full blue line contour) and $95\%$ in orange (with dashed contours) for the present fit. The gray bands represent the Soffer bound at NLO, here using JAM and CT18.}
\label{fig:12sigm_UD}
\end{figure}
\begin{figure}
\begin{center}
		\includegraphics[scale= .53]{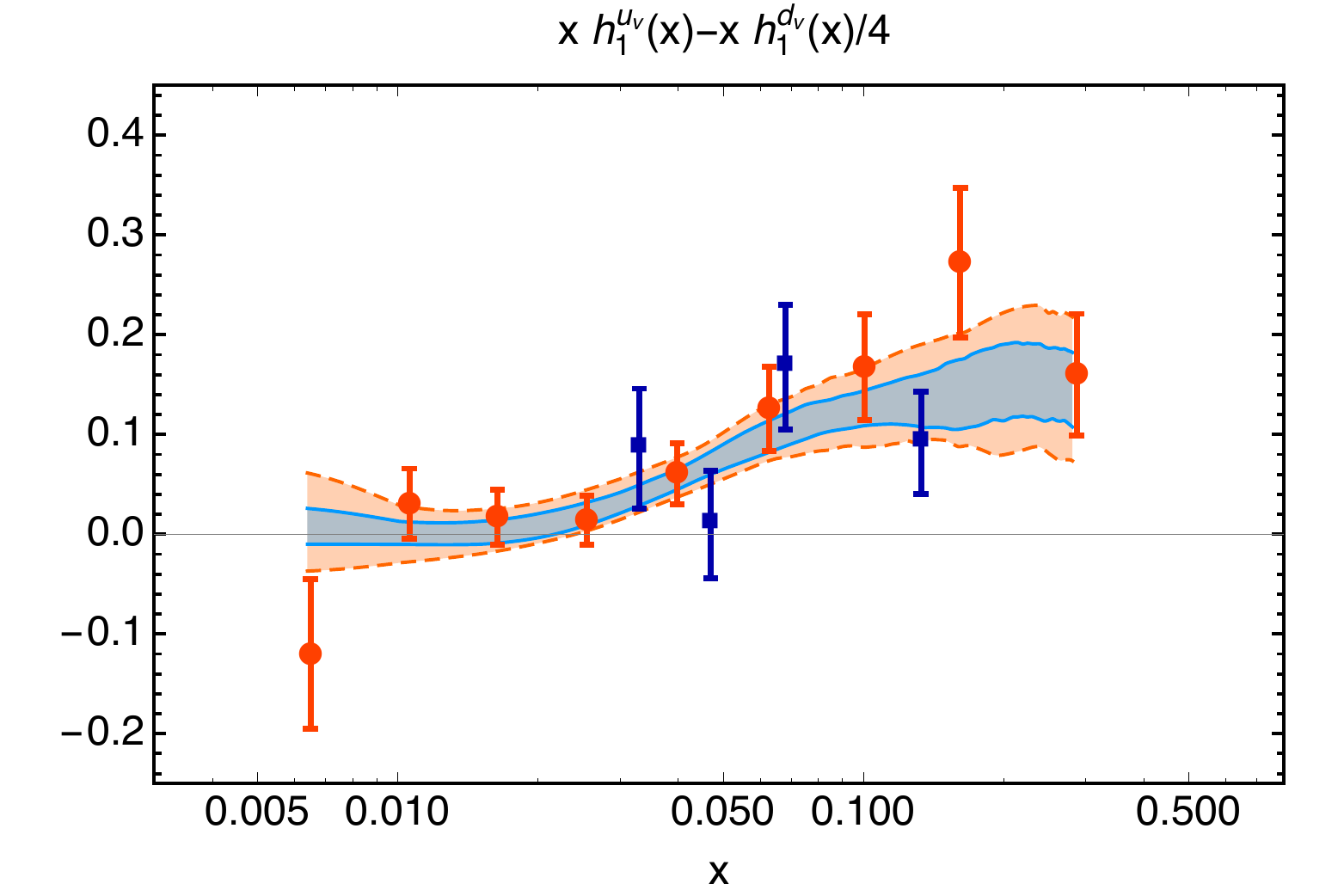}
		\includegraphics[scale= .53]{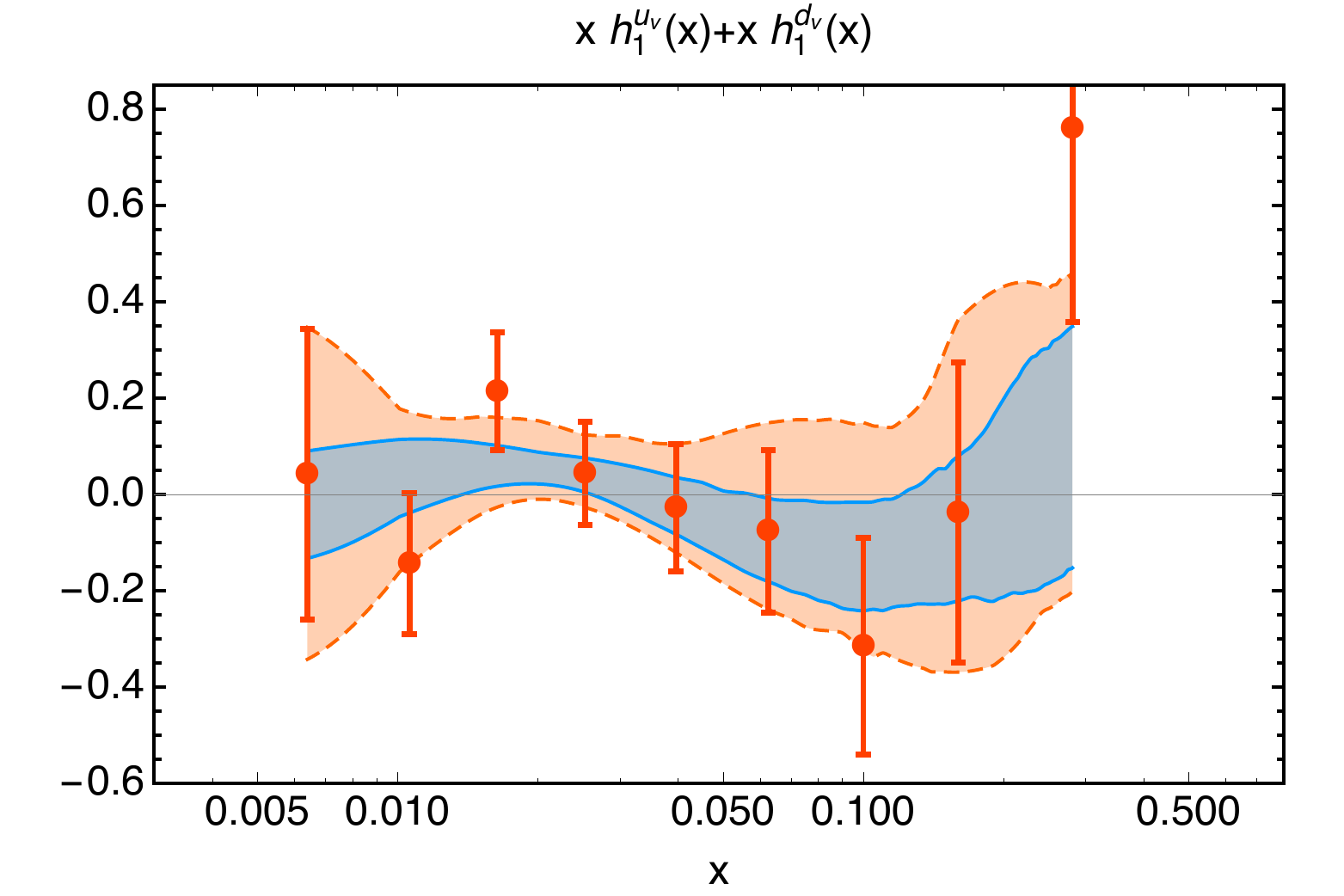}
\end{center}
\caption{Combinations of transversities for the proton (left) and the deuteron (right) compared to the global $68\%$ in blue (with full blue line contour) and $95\%$ in orange (with dashed contours) for the present fit compared to the extracted data~\cite{Radici:2015mwa}. The dark blue square come from HERMES data while the dark orange dots are extracted from COMPASS data.}
\label{fig:12sigm_data}
\end{figure}

\section{Results}
\label{sec:res}

\subsection{Transversity PDF}

Both steps of the minimization procedure are carried out in Python. The convergence for both the main $\chi^2$ minimization Eq.~(\ref{eq:chi2main}) and the optional Lagrange multipliers method are fast.  Twenty five percents of the replicas converge inside the established bound at the first minimization. The $75\%$ left is hence constrained by the method of Lagrange multipliers. For the higher-order polynomials, less than 15 additional iterations are necessary while the two lower-order polynomials require about 30-40 extra iterations. Although the constraints are imposed for the down distribution only, their fulfillment affects directly the parameterization of the up. The balance between $u_v$ and $d_v$ is clearly passed on through the objective function. Notice that not all 4 constraints are active for each replica $r$.

In Fig.~\ref{fig:UD_44}, we show the envelopes for the obtained valence transversities at $68\%$ CL, respectively, for the up and the down distributions. Following the color code introduced in Fig.~\ref{fig:BPUD_44}, it can be observed that the higher order polynomials --$n=30,40$ in  yellow and green-- allow a larger error bar at smaller values of $x$. On the other hand, the lower order polynomials --$n=10,20$ in red and purple-- are confined in the mid- to large-$x$ region. All 4 equally contribute at valence values of the Bjorken variable. The final envelope of $xh_1^{q_v}(x,\langle 5 \mbox{GeV}^2\rangle)$ is built using the four versions of the functional form together. It is shown at $68\%$ and $95\%$ CL in Fig.~\ref{fig:12sigm_UD}. For consistency, we also show the Soffer bound at NLO using the helicity PDF of JAM15~\cite{Sato:2016tuz} and the unpolarized PDF of CT18~\cite{Hou:2019qau}. Were there central values for the obtained transversities, they would be enclosed inside the Soffer bound. In that sense, our result is similar to the first Hessian approach of Ref.~\cite{Bacchetta:2012ty}. 

The small-$x$ behavior predicted in Ref.~\cite{Kovchegov:2018zeq}  is  remarkably  fulfilled for the up transversity PDF up to, at least, $x=0.1$ ; no clear conclusion can be drawn for the down  above  $x\sim 0.03$.

\begin{figure}
\begin{center}
		\includegraphics[scale= .48]{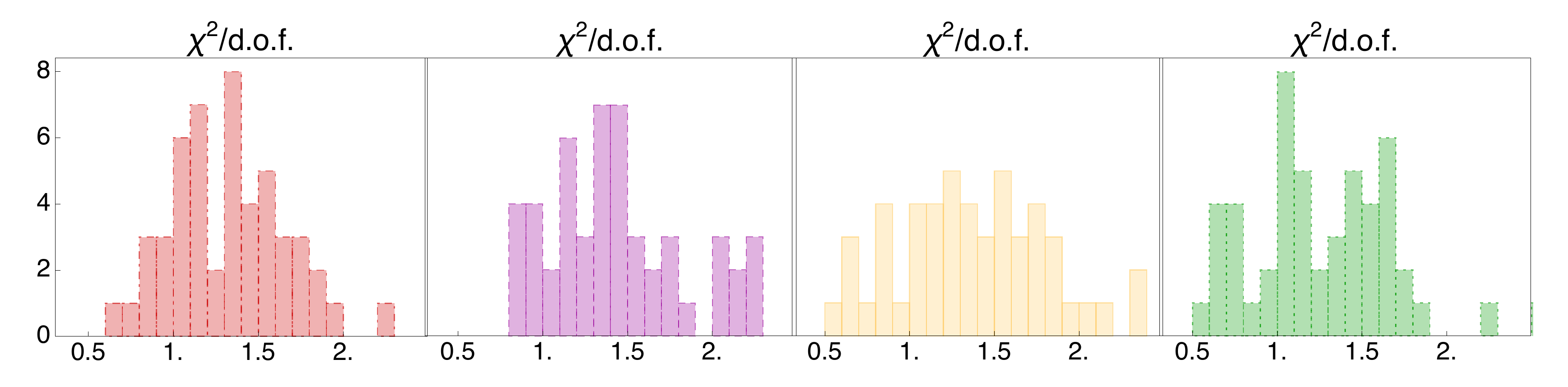}
\end{center}
\caption{Histograms of chi-square per degree of freedom for each parametrization. The color code represent, as above,  the four different degree of the Bernstein polynomial of the functional form, respectively, $n=10, 20, 30, 40$ in red, purple, yellow and green. }
\label{fig:chi2}
\end{figure}

In Fig.~\ref{fig:12sigm_data}, the $68\%$ and $95\%$ CL proton and deuteron combinations are depicted and compared to the point-by-point extraction of Eqs.~(\ref{eq:prot}-\ref{eq:deut}). The final $\chi^2/$d.o.f. is evaluated against those extracted data. The number of degrees of freedom here is $(\sum_j\, w_j)\times N_d+N_c-(n_{\kappa_{u}}+n_{\kappa_{d}})$. The average value for the total $\chi^2/$d.o.f. is $1.35$. 
Each functional form contributes to average values of $\chi_{n=10}^2/$d.o.f.$=1.32$, $\chi_{n=20}^2/$d.o.f.$=1.44$, $\chi_{n=30}^2/$d.o.f.$=1.37$, $\chi_{n=40}^2/$d.o.f.$=1.26$.
The four corresponding  histograms are shown in Fig.~\ref{fig:chi2}. 
\\

The analysis has been carried out using the MSTW08LO parametrization. We have repeated the fitting procedure using the  CT09MC1 set for the unpolarized PDF~\cite{Lai:2009ne}, it is a LO set fitted on real data and NLO pseudo data with $\alpha_s$ at 1-loop. There is no qualitative difference in the  final transversity PDF.  We notice a slightly smaller average value  for the isovector tensor charge, quantity that will be defined here below. 

\begin{figure}
\begin{center}
		\includegraphics[scale= .45]{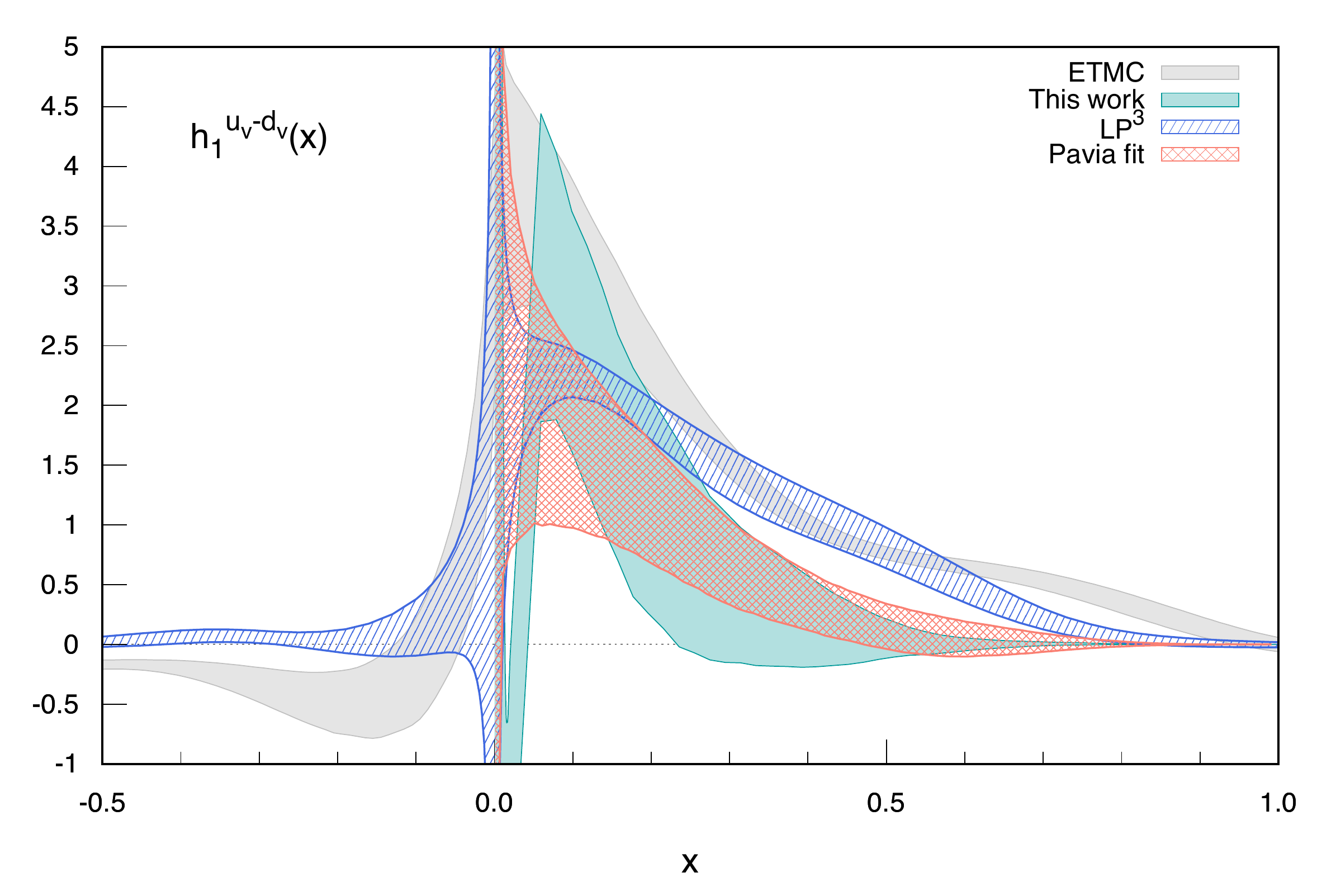}
\end{center}
\caption{Comparison with the isovector combination of the transversity  PDF of this work, in cyan, with the result at $4$ GeV$^2$ obtained through the global fit of Ref.~ \cite{Radici:2018iag}, in meshed pink, both at 90 \% CL. These collinear extractions are compared to  lattice QCD evaluations from the ETM Collaboration~\cite{Alexandrou:2018eet}, in gray, and the LP$^3$ Collaboration~\cite{Liu:2018hxv}, in dashed blue. Lattice results correspond to a scale of $4$ GeV$^2$.}
\label{fig:compLatt}
\end{figure}

Compared to previous extractions, we might comment that our result for the down transversity presents a smaller error band in the data region --as our functional form was built on that purpose. It is achieved at the expense of a slightly wider band for the up distribution in that region. The error band increases for small- and large-$x$ values. In this respect, our down distribution differs from that of Refs.~\cite{Radici:2015mwa,Radici:2018iag} and the less flexible versions of the parameterization of Ref.~\cite{Bacchetta:2012ty}, all of them being dihadron-based extractions.   The same comment is in order for comparisons with single-hadron semi-inclusive DIS~\cite{Anselmino:2013vqa,Kang:2015msa}.
The combination $h_1^{u_v}-h_1^{d_v}$ of the two available collinear extractions is compared in Fig.~\ref{fig:compLatt}. The small-$x$ behavior differs and the error band associated to the present analysis is wider in the valence region.

We also compare, in Fig.~\ref{fig:compLatt}, our results to the lattice QCD evaluation of the isovector transversity PDF~\cite{Alexandrou:2018eet,Liu:2018hxv}. The order of magnitude as well as mid-$x$ behavior are in a reasonable agreement. The lower-$x$ region of our result is more structured, as dictated by the data. Both lattice evaluations exceed the phenomenological determinations at large-$x$ --region in which the parameterization of the fits are strongly bound by the positivity limits.

As explained in Appendix~\ref{sec:form}, our analysis of the transversity PDF through Eqs.~(\ref{eq:prot}, \ref{eq:deut}) has been carried out  at an average value of $Q^2=\langle 5 $GeV$^2\rangle$. 
To that regard we have further performed our analysis considering QCD evolution only for the Fragmentation Functions.  We find a 15\% increase in the $\chi^2$ when fixing the scale of the unpolarized PDFs in the asymmetry to 5 GeV$^2$. The comparison with the isovector transversity from lattice QCD is not substantially improved.

Finally, we notice that the PDF$+$Lattice result for the isovector combination~\cite{Lin:2017stx} is systematically higher than the present results for the whole support in $x$, except for $x\to 0$. 
\\

\subsection{Tensor Charge}

\begin{figure}[t]
\begin{center}
		\includegraphics[scale= .53]{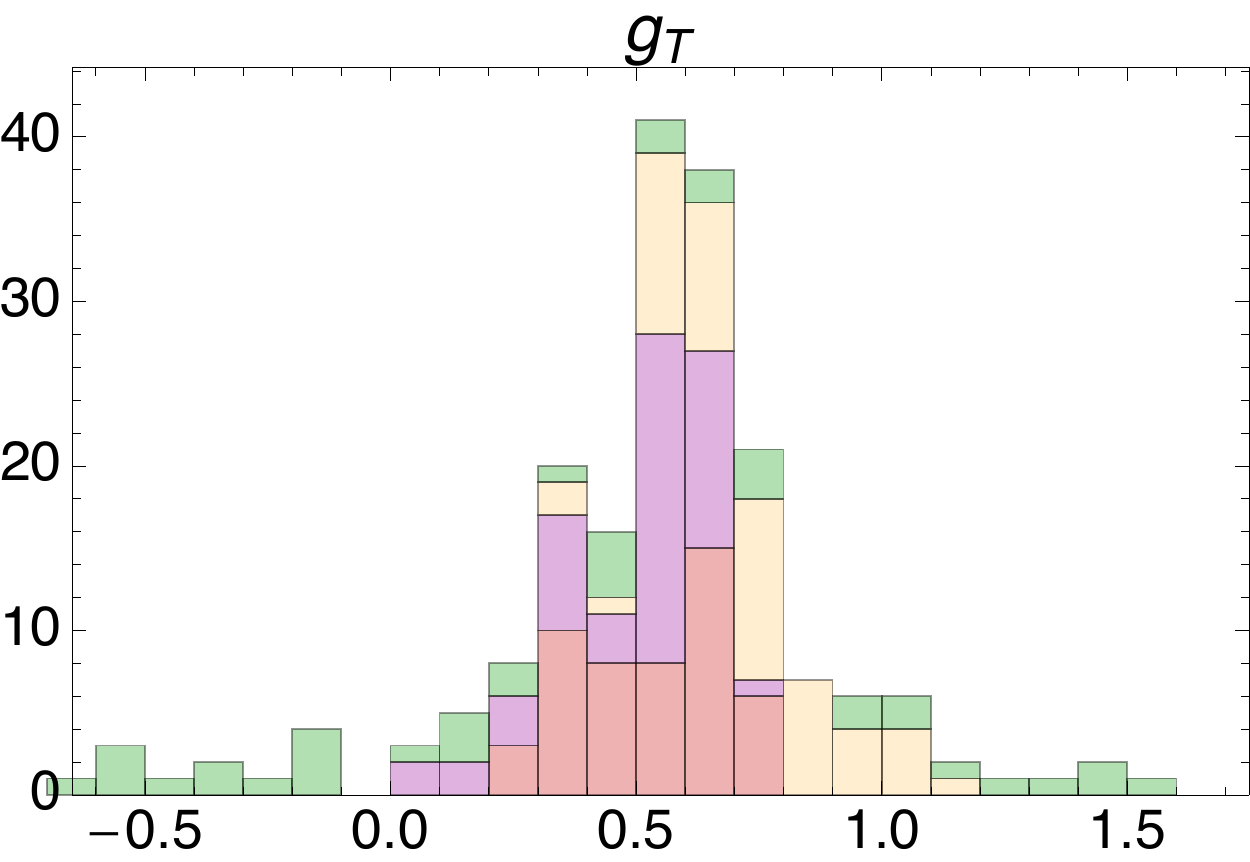}
		\includegraphics[scale= .53]{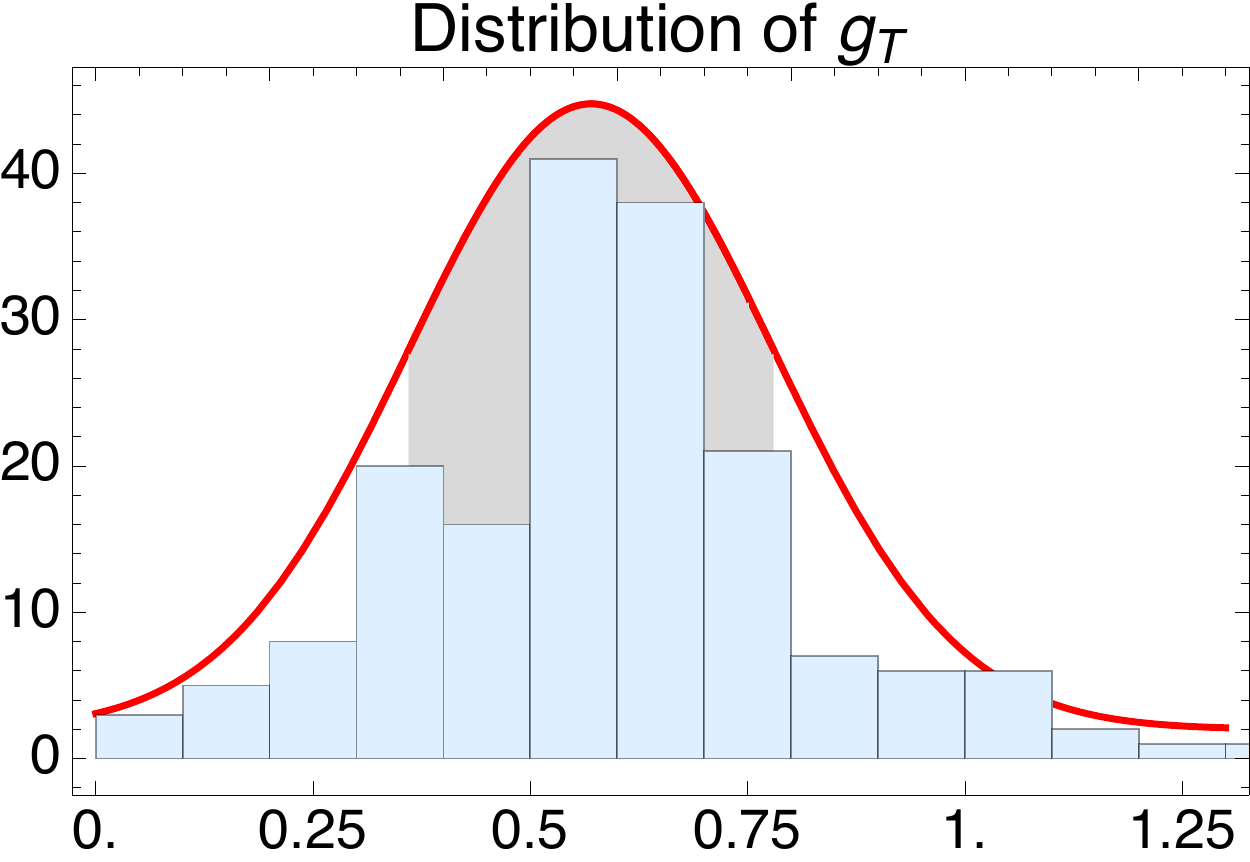}
\end{center}
\caption{Stacked histogram of isovector tensor charge. The color code represent, as above,  the four different degrees of the Bernstein polynomial of the functional form, respectively, $n=10, 20, 30, 40$ in red, purple, yellow and green. Notice that the various contributions are represented on a single bar, {\it i.e. stacked}. On the {\it r.h.s.}, we can see that $g_T$ is  gaussianly distributed, as shown in red with the corresponding $1\sigma$ error in gray.}
\label{fig:gT}
\end{figure}

We next evaluate the first Mellin moment of the transversity PDF to get the tensor charge
\begin{eqnarray}
\delta q_v \left(5 \mbox{GeV}^2\right) &= &\int_0^1 dx \, h_1^{q_v} \left(x;  5 \mbox{GeV}^2\right) \; . 
\label{eq:deltaq}
\end{eqnarray}
The distributions correspond best to skewed distributions  and the obtained values are
\begin{eqnarray}
\delta u_v \left( 5 \mbox{GeV}^2\right) &= & 0.28\substack{+0.17 \\ -0.20}\quad,\nonumber\\
\delta d_v \left( 5 \mbox{GeV}^2\right) &= & -0.40\substack{+0.41 \\ -0.31}\quad,
\end{eqnarray}
to $1\sigma$. The corresponding $2\sigma$ uncertainties are, for the up,~$(-0.42,0.31)$ and, for the down,~$(-0.57,0.87)$.
In particular, we are interested in the isovector combination, $g_T(Q^2)$, as it is of particular interest for, {\it e.g.}, beta decay observables~\cite{Bhattacharya:2011qm,Courtoy:2015haa}. We show the corresponding stacked histogram in Fig.~\ref{fig:gT}. The Gaussian distribution showed on the {\it r.h.s.} of Fig.~\ref{fig:gT} is given by 
\begin{eqnarray}
g_T \left( 5 \mbox{GeV}^2\right) &= & 0.57\pm 0.21\quad,
\end{eqnarray}
at $1\sigma$ ; the $2\sigma$ error is $0.42$.
The truncated tensor charge is found to be $g_T^{\urcorner}( 5 $GeV$^2)~=~0.48\substack{+0.13 \\ -0.11}$, corresponding to a skewed distribution. 

The trend observed above leads to discernible consequences here: the more flexible the parameterization, the wider the distribution of the tensor charge.  Our result is compatible with lattice QCD evaluations of the isovector tensor charge --the latest results are $g_T=0.926 (32)$ for the ETM Collaboration~\cite{Alexandrou:2019brg}, $g_T = 0.972 (41)$~\cite{Hasan:2019noy} and $g_T = 0.965 (38) (+13 -41)$~\cite{Harris:2019bih}. It is at the same time  wide enough to encompass all of the other phenomenological determinations. 
As for the separate up and down contributions, {\it e.g.}, the ETM Collaboration reports $\delta u=0.716(28)$ and $\delta d=-0.210(11)$. The tensor charge for the up quark differs from our result as well as most phenomenologically determined tensor charges. The relaxation of the Soffer bound in the range of the data did not ease that discrepancy, as also discussed in Ref.~\cite{DAlesio:2020vtw}.

In a region where  scarce to no data are available, the choice of functional form generates a crucial uncertainty.
 The role of the small-$x$ region, where no strong expressions of first principle bounds nor data are available, matters. The addition of proton-proton collision data~\cite{Adamczyk:2015hri}, used in the first global fit of the transversity~\cite{Radici:2018iag},  does unfortunately not increase the coverage at lower $x$ values --but it certainly does in $Q^2$. In that sense, data from an Electron Ion Collider would be extremely helpful, be it for a qualitative improvement over a quantitative reduction of the uncertainty. Would the tensor charge become relevant in view of search for New Physics, the appropriate observables would not necessarily linearly depend on the uncertainty of the tensor charge, as shown in the case of beta decay in Ref.~\cite{Courtoy:2015haa}.

\section{Conclusions}
\label{sec:conclu}

The transversity Parton Distribution Function is the least known of the three leading-twist PDFs. Its phenomenological  determination has been made possible thanks to independent data for its partners in semi-inclusive DIS, the chiral-odd fragmentation functions. In particular, the formalism developed around dihadron Semi-Inclusive DIS allows for a collinear extraction of combinations of valence transversity PDFs~\cite{Radici:2001na}. It requires the knowledge on the Dihadron Fragmentation Functions that have been studied in Ref.~\cite{Courtoy:2012ry,Radici:2015mwa}. Since its first point-by-point extraction from dihadron semi-inclusive DIS~\cite{Bacchetta:2011ip},  enormous efforts  towards the extraction of the transversity PDF,  accessible in processes involving fragmentation to a pion-pair~\cite{Radici:2018iag}, have been made.

In this paper we have presented a constrained fit of the valence transversity PDFs from dihadron semi-inclusive DIS data. The adopted methodology is characterized by the flexibility to accommodate constraints from the fitting procedure. It consists in three steps. First we have examined the positivity constraints on the transversity distribution: a {\it prior} from the expression of the bound combined with the data is introduced as a theoretical uncertainty on the PDF that is then implemented through a reweighting of the data. 
Second we have chosen  flexible functional forms --free from ad hoc expressions of the Soffer bound-- to enhance the information obtained from the data and first principles. It results in an improved treatment of the PDF uncertainties. Finally, we have guided the valence down transversity PDF to follow a fall-off behavior at  $x\to 1$  through the method of Lagrange multipliers, forcing the set of parameters to observe the theoretical constraints.

The obtained $68\%$ and $95\%$ CL envelopes for the valence transversity distribution functions fulfill the expected small-$x$ behavior, the up distribution fully realizes the Soffer bound and the edges of the down envelope reflect the relaxation of the bound combined with the lack of data in the large-$x$ region.
Our results globally show wider error bands outside the data range with respect to the error band inside that range than found in previous extractions.  That trend, allowed by the --absence of-- data, translates into a wide distribution of the tensor charge yet with a more comfortable extrapolation in the whole support.  The resulting isovector tensor charge is in agreement with the determinations of lattice QCD. We believe that this exercise supports the idea that the choice of functional form contributes to the global error of the PDF determination, especially in kinematical regions with limited data coverage.

At SIDIS scale the relative effect of DGLAP is small compared to the accurracy of the obtained PDFs.  The effect of QCD evolution will become important when including data from proton-proton collision from RHIC for which the role of NLO corrections might become important, as discussed in Ref.~\cite{Radici:2018iag}.  Our methodology could be applied, as a natural extension of this work, to the extended set of data including the aforementioned proton-proton data and the inclusion of the corresponding DGLAP routine. More semi-inclusive DIS data in the valence region are expected from CLAS12 and SoLID at JLab; data from the future Electron Ion Collider would improve the uncertainty in the  low-$x$ region.

\section*{Acknowledgments}
The authors thank S.~Fromenteau, P.~Nadolsky  and M.~Vargas-Maga\~na for illuminating discussions. The work of A.C. and R.F.-H. is supported by UNAM through the PIIF project Perspectivas en F\'isica de Part\'iculas y Astropart\'iculas,  DGAPA-PAPIIT IA102418 and IA101720, and the German--Mexican research collaboration grant SP 778/4--1 (DFG) and 278017 (CONACyT). R.F.-H. is also supported in part by CONACyT  project 252167--F.
J.B. has been funded by Beca Alianza del Pac\'ifico 2018-2 and UdeI-FC of the Universidad Nacional de Ingenier\'ia, Peru.

\appendix
\section{Transverse spin asymmetries in semi-inclsuive deeply inelastic scattering}
\label{sec:form}

We focus on assessing  the transversity PDF  through the measurement of the single-spin asymmetry in dihadron semi-inclusive production off a proton/deuteron target process,
\begin{equation}
l(k) + N(P) \to l(k') + \left(H_1(P_1) + H_2(P_2)\right) + X \quad.
\end{equation}
The initial-state (unpolarized) lepton  $l$ with 4-momentum $k$ and the outgoing lepton $k'$ is scattered on the transversely polarized target $N$, which can be a proton or deuteron inside a nuclear target, {\it e.g.} hydrogen at HERMES~\cite{Airapetian:2008sk} or $^6$LiD (deuteron) and NH$_3$ (proton)  at COMPASS~\cite{Adolph:2012nw}. In total, there are $N_d=22$ data points.
The produced hadrons are labelled $H_1$ and $H_2$,  their 
invariant mass squared  is $P_h^2 = M_h^2$, with $P_h$, total momentum of the pair. Their relative momentum is defined as $R = (P_1-P_2)/2$. The 
momentum transferred to the nucleon target is $q = k - k'$. The usual DIS invariants are defined, $x=Q^2/(2 P\cdot q)$, $y=P\cdot q/P\cdot k$, and $z=z_1+z_2=P\cdot P_h/ (P\cdot q)$. We refer to Ref.~\cite{Radici:2015mwa} and references therein for further details.

In this process, the chiral-odd  PDF couples to a similarly chiral-odd DiFF --understood as the distribution of a hadron pair whose orbital angular motion through its relative transverse momentum, ${\bf R}_T$, originates from the transversely polarized fragmenting quark~\cite{Radici:2001na}.
The transverse momentum ${\bf R}_T$ can be expressed in terms of the invariant mass of the pair and the polar angle $\theta$ of the one pion relative to $P_h$ in their center-of-mass.
\\

The state-of-the-art formalism for the transversity in a collinear framework has been set, to leading-order, in Ref.~\cite{Bacchetta:2002ux}.
In the limit $M_h^2<<Q^2$ and after having selected the dominant contribution through Partial Wave Expansion in the polar angle  $\theta$, the relevant single-spin asymmetry  reads
\begin{equation}
A_{\mathrm{SSA}} (x, z, M_h; Q) = - \frac{B(y)}{A(y)} \,\frac{|\bf{R} |}{M_h} \, 
\frac{ \sum_q\, e_q^2\, h_1^q(x; Q^2)\, H_1^{\open\, q}(z, M_h; Q^2)    } 
        { \sum_q\, e_q^2\, f_1^q(x; Q^2)\, D_{1}^q (z, M_h; Q^2) } \quad,
\label{e:ssa}
\end{equation} 
where $e_q$ is the fractional charge of a parton with flavor $q$, $A(y) = 1-y+y^2/2, \;B(y) = 1-y$. The $D_1^q$ is the DiFF describing the hadronization of an  unpolarized parton with flavor $q$ into an unpolarized hadron pair. The $H_1^{\open\, q}\equiv H_{1,sp}^{\open\, q}$ is a chiral-odd DiFF describing the correlation between the transverse polarization of the fragmenting parton with flavor $q$ and the azimuthal orientation of the plane containing the momenta of the detected hadron pair. Both have been determined from Belle's data and corresponding PYTHIA-generated multiplicities~\cite{Courtoy:2012ry,Radici:2015mwa}.

As the PDFs only depend on $x$, we will only use the $x$-projected part of $A_{\mathrm{SSA}}(x,z,M_h; Q)$, obtained by integrating out  $z$ and $M_h$ over the corresponding kinematics of the experiment. The relevant combinations  for the proton 
target are~\cite{Bacchetta:2012ty} 
\begin{eqnarray} 
x\, h_1^{p}\left(x; \langle Q^2\rangle\right) &\equiv &x \, h_1^{u_v}\left(x; \langle Q^2\rangle\right) -  \frac{1}{4}\, x h_1^{d_v}\left(x; \langle Q^2\rangle\right) \nonumber\\
\nonumber\\
&=& -\frac{ A^p_{\text{SSA}} (x; Q^2)  }{n_u^{\uparrow}(Q^2)} \, \frac{A(y)}{B(y)}\, \frac{9}{4}\, 
\sum_{q=u,d,s} \, e_q^2 \, n_q (Q^2)\, x f_1^{q+\bar{q}}(x; Q^2)\; ,
\label{eq:prot}  
\end{eqnarray} 
and for the deuteron target are 
\begin{eqnarray} 
 x\, h_1^{D} \left (x; \langle Q^2\rangle\right) &\equiv& x \, h_1^{u_v}\left(x; \langle Q^2\rangle\right)+ x h_1^{d_v}\left(x; \langle Q^2\rangle\right)  \nonumber \\
 \nonumber\\
 &=&- \frac{A^D_{\text{SSA}}(x; Q^2)}{n_u^{\uparrow}(Q^2)} \, 3  \, 
 \sum_{q=u,d,s} \, \left[ e_q^2 \, n_q (Q^2) + e^2_{\tilde{q}} \, n_{\tilde{q}} (Q^2) \right] \, x f_1^{q+\bar{q}}(x; Q^2)\; ,
\label{eq:deut}
\end{eqnarray} 
where $h_1^{q_v}$ is the valence combination, {\it i.e.} $ h_1^q - h_1^{\bar{q}}$, and $f_1^{q+\bar{q}} \equiv f_1^q + f_1^{\bar{q}}$ and $\tilde{q}= d, u, s$ if $q = u, d, s$, respectively. Unless explicitly stated, we use the MSTW08LO parametrization~\cite{Martin:2009iq} to evaluate the unpolarized PDFs. The quantities $n_q (Q^2)$ and  $n_u^{\uparrow}(Q^2)$ are, respectively,  the unpolarized DiFF for a flavor $q$ and the chiral-odd DiFF for $u$  that are integrated over $z$ and  $M_h$ in the given experiment. .

We have not taken into account the $Q^2$ dependence of the transversity PDF --on the {\it l.h.s.} of Eqs.~(\ref{eq:prot}-\ref{eq:deut})-- but rather considered its parametrization as given at the average scale $\langle Q^2\rangle=5 $GeV$^2$ of the data --justified by the relatively narrow span of values of $Q$.  The dependence on the hard scale is kept in all the quantities on the {\it r.h.s} of Eqs.~(\ref{eq:prot}-\ref{eq:deut}) for the purpose of assessing properly the statistical weight of the Soffer bound on the point-by-point extraction. On the other hand, we do not expect exact cancellations between numerator and denominator~\cite{Bacchetta:2011ip}, in that sense our approximation is slightly different from that adopted in Ref.~\cite{Lin:2017stx}. Consequences of this approximation are discussed in Section~\ref{sec:res}. 

In this paper, we use the relevant single-spin asymmetry measured at HERMES~\cite{Airapetian:2008sk} and COMPASS for identified hadron pairs~\cite{Braun:2015baa,Braunthesis}. These data sets coincide with those used in Ref.~\cite{Radici:2015mwa}. The  kinematical ranges are $1.2<Q^2 <31.5$ GeV$^2$, $0.0064<x<0.2871$ --with increasing values of $Q^2$ for increasing values of $x$-- and the DiFF variables are $2m_{\pi}<M_h\lesssim 1.3 $GeV and $0.2\lesssim z<0.9$. A point-by-point analysis of dihadron and single-hadron SIDIS COMPASS data has been proposed in Ref.~\cite{Martin:2014wua} to confirm their overall compatibility.


\bibliographystyle{hunsrt}
\bibliography{bib_transv}

\end{document}